\documentclass[
twocolumn,
aps,
prb,
reprint,
groupedaddress
]{revtex4-1}

\usepackage{graphicx}
\usepackage{amssymb,amsfonts,amsmath}
\usepackage[utf8x]{inputenc}

\newcommand{\Ilaser}{I_{\mathrm{L}}}
\newcommand{\Llaser}{\lambda_{\mathrm{L}}}
\newcommand{\Ton}{\tau_{\mathrm{on}}}
\newcommand{\Toff}{\tau_{\mathrm{off}}}

\begin{document}

\title{Single spin stochastic optical reconstruction microscopy}

\author{Matthias Pfender}
\altaffiliation{these authors contributed equally}
\author{Nabeel Aslam}
\altaffiliation{these authors contributed equally}
\author{Gerald Waldherr}
\author{Philipp Neumann}
\email{p.neumann@physik.uni-stuttgart.de}
\author{J\"org Wrachtrup}
\affiliation{3rd Institute of Physics, Research Center SCoPE and IQST, University of Stuttgart, 70550 Stuttgart, Germany}

\begin{abstract}
We experimentally demonstrate precision addressing of single quantum emitters by combined optical microscopy and spin resonance techniques.
To this end we utilize nitrogen-vacancy (NV) color centers in diamond confined within a few ten nanometers as individually resolvable quantum systems.
By developing a stochastic optical reconstruction microscopy (STORM) technique for NV centers we are able to simultaneously perform sub diffraction-limit imaging and optically detected spin resonance (ODMR) measurements on NV spins.
This allows the assignment of spin resonance spectra to individual NV center locations with nanometer scale resolution and thus further improves spatial discrimination.
For example, we resolved formerly indistinguishable emitters by their spectra.
Furthermore, ODMR spectra contain metrology information allowing for sub diffraction-limit sensing of, for instance, magnetic or electric fields with inherently parallel data acquisition.
As an example, we have detected nuclear spins with nanometer scale precision.
Finally, we give prospects of how this technique can evolve into a fully parallel quantum sensor for nanometer resolution imaging of delocalized quantum correlations.
\end{abstract}

\maketitle

\section{Introduction}
Stochastic reconstruction microscopy (STORM) techniques have lead to a wealth of application in fluorescence imaging \cite{huang_three-dimensional_2008,huang_whole-cell_2008,egner_fluorescence_2007}, for example few ten nanometers three-dimensional spatial resolution has been achieved in cellular imaging.
So far, STORM fluorophores have been used as markers to achieve nanoscale microscopy of specific targets \cite{hild_quantum_2008}.
Here, we present a novel spin-based approach which promises to combine sub diffraction-limit imaging via STORM and simultaneous sensing of various physical quantities.
\begin{figure*}[t]
	\begin{center}
	\centerline{\includegraphics[width=1.0\textwidth]{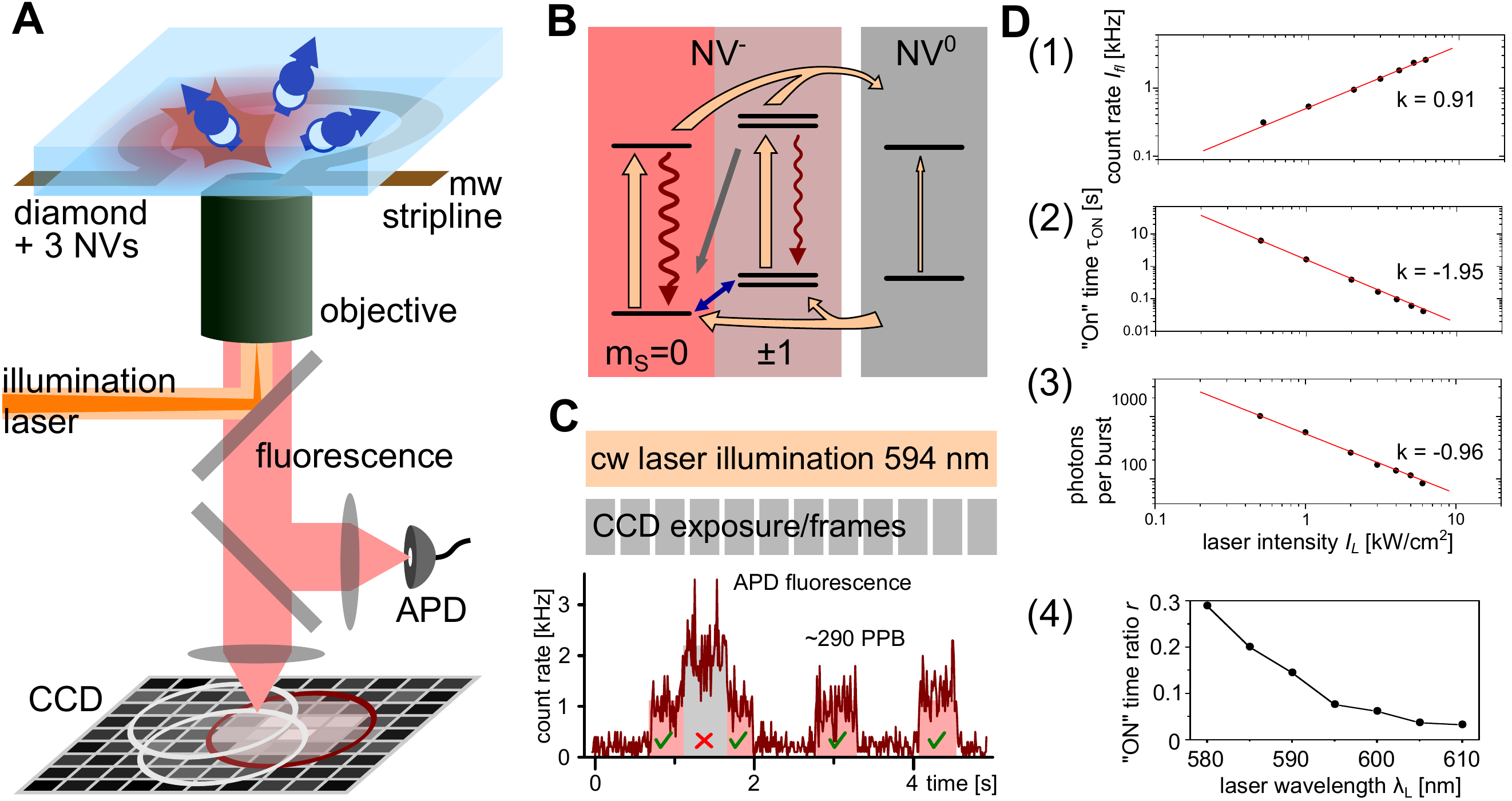}}
  \caption{\textbf{The NV center in diamond and the experimental setup.}
	(\textit{A}) Experimental setup (see text).
	In this illustration one of three NV centers in the confocal spot is in the negative charge state (``On'') the others are neutral (``Off'').
	An exemplary CCD image with marked PSFs is sketched.
	(\textit{B}) Simplified energy level scheme for neutral and negatively charged NV center.
	Laser induced rates are shown as orange arrows and luminescence is shown as wavy lines.
	The line widths symbolize strengths of the transitions.
	The NV spin polarization rate is sketched as gray arrow and mw induced spin transitions are shown as blue double arrow.
	(\textit{C})
	Upper part displays the STORM measurement scheme for NV centers (see text).
	Laser illumination 
	is applied continuously and	CCD imaging is performed with a fixed frame rate.
	The lower part shows an exemplary fluorescence time trace recorded with an APD.
	Lowest count level indicates that all NV defects are in the ``Off'' state.
	Two distinguishable higher fluorescence levels correspond to one and two NV centers in the ``On'' state.
	(\textit{D}) (1) Laser intensity dependence of single NV center fluorescence count rates $\Gamma$.
	(2) Laser intensity dependence of NV$^{-}$ charge state lifetime $\tau_{\mathrm{on}}$ (i.e. ``On'' time).
	(3) Laser intensity dependence of average photons per burst $n$.
	(4) The NV$^{-}$ charge state fraction (i.e. $r=\Ton/\left(\Ton + \Toff\right)$) for excitation wavelengths between 580 nm and 610 nm.%
	\label{fig:setup}}
	\end{center}
\end{figure*}
\\
\indent
As a prominent multipurpose probe and
and highly photostable single emitter
we use the nitrogen-vacancy spin defect in diamond.
It can be applied for nanometer scale scanning magnetometry \cite{balasubramanian_nanoscale_2008,maze_nanoscale_2008,grinolds_nanoscale_2013,rondin_stray-field_2013} as well as magnetic imaging \cite{steinert_high_2010,steinert_magnetic_2013,le_sage_optical_2013,maertz_vector_2010,pham_magnetic_2011,hall_high_2012} (e.g. for imaging spin distributions, magnetic particles or organisms, or device intrinsic fields), the measurement of electric fields and diamond lattice strain \cite{dolde_electric-field_2011,dolde_nanoscale_2014,macquarrie_mechanical_2013,michl_perfect_2014-1,doherty_electronic_2014} (e.g. for imaging elementary charges or charge distributions, or for imaging strain fields induced by mechanical action on the diamond surface).
Very recently precise temperature measurements \cite{toyli_fluorescence_2013-2,neumann_high-precision_2013} even in living cells \cite{kucsko_nanometre-scale_2013} have been demonstrated.
\\
\indent
During the last decades a variety of methods have been invented to circumvent the diffraction limit in farfield optical microscopy. 
One approach reduces the spatial region within a laser focus from which optical response of a single emitter is possible by exploiting optical nonlinearities.
Examples are stimulated emission depletion (STED) and ground state depletion (GSD) microscopy \cite{hell_breaking_1994,hell_ground-state-depletion_1995}.
Another approach tailors the timing of optical response of several emitters from within a diffraction limited spot to distinguish them in the time domain.
One example is stochastic optical reconstruction microscopy \cite{betzig_imaging_2006,hess_ultra-high_2006,rust_sub-diffraction-limit_2006}.
This latter technique is intrinsically parallel as it utilizes a CCD array for imaging and is therefore particularly suited for high throughput imaging.
\\
\indent
STED and GSD microscopy, which are both scanning techniques, have been recently implemented for NV centers in diamond \cite{rittweger_sted_2009,rittweger_far-field_2009,han_metastable_2010} with resolutions down to a few nanometers \cite{wildanger_solid_2012}. In addition, localization-based superresolution microscopy has been shown with NV centers in nano-diamonds \cite{gu_super-resolving_2013}.
\\
\indent
Here, we experimentally demonstrate STORM for NV centers in diamond as a new optical super-resolution technique
with wide-field parallel image acquisition for NV centers in bulk diamond.
Our technique is based on recently gained profound knowledge about statistical charge state switching of single NV centers \cite{aslam_photo-induced_2013} and its scalability relies on the homogeneity of this charge state dynamics for NV centers in bulk diamond.
Furthermore, we combine optical superresolution microscopy with high spectral resolution optically detected magnetic resonance (ODMR).
On the one hand, we use the latter technique to assign magnetic resonance data to nanometer scale locations, which is important for qubit or metrology applications \cite{dolde_room-temperature_2013,steinert_high_2010,steinert_magnetic_2013,le_sage_optical_2013}.
On the other hand different magnetic resonance fingerprints of closely spaced NV centers are used to further increase the already obtained superresolution, as demonstrated in \cite{dolde_room-temperature_2013,chen_wide-field_2013}, which is important for emitter localization in imaging applications.

\section{Results}
\subsection{Relevant key features of NV centers in diamond}
The negatively charged NV center in diamond is an optically active emitter with an electronic spin in its ground state (see fig.~\ref{fig:setup}\textit{B}) with favorable coherence properties \cite{balasubramanian_ultralong_2009,mizuochi_coherence_2009}.
The NV's special properties allow for optical detection of single centers and the optical initialization and readout of its electronic spin \cite{gruber_scanning_1997}.
The latter is used for metrology \cite{balasubramanian_nanoscale_2008,acosta_temperature_2010,toyli_measurement_2012,neumann_high-precision_2013,dolde_electric-field_2011} and quantum information processing (QIP) \cite{dutt_quantum_2007,togan_quantum_2010,dolde_room-temperature_2013,bernien_heralded_2013-2,waldherr_quantum_2014} applications.
More specifically, the NV center in diamond is a point defect in the diamond lattice consisting of a substitutional nitrogen atom next to a carbon vacancy. It appears mainly in two different charge states, NV$^-$ and the neutral NV$^0$ \cite{aslam_photo-induced_2013} (see fig.~\ref{fig:setup}\textit{B}).
Upon illumination in a wide spectral range ($\approx 500 \ldots 637\,$nm) fluorescence indicating the charge state can be invoked. 
The spectral excitation windows as well as the fluorescence spectra for NV$^{0}$ are slightly blue shifted with respect to NV$^{-}$, exhibiting a zero-phonon line (ZPL) at $\approx 575\,$nm and $\approx 637\,$nm respectively accompanied by phonon sidebands \cite{aslam_photo-induced_2013}.
\\
\indent
Recently, frequent switching between NV$^-$ and NV$^0$ charge states has been demonstrated, and can be detected via the presence and absence of fluorescence respectively \cite{waldherr_dark_2011,aslam_photo-induced_2013}.
This stochastic fluorescence switching (i.e. visible bursts of fluorescence from single NV centers; see fig.~\ref{fig:setup}\textit{C}) is exploited in our work for STORM.
By adjusting illumination intensity $\Ilaser$ and wavelength $\Llaser$ we can tune parameters such as fluorescence burst length $\Ton$, photons per burst $n$ and the ``On'' fraction $r=\Ton/\left(\Ton+\Toff\right)$ of the emitter (see fig.~\ref{fig:setup}\textit{C,D} and Methods).
As an example, for a single NV center and $\Ilaser \sim 1\,$kW/cm$^2$ of $\Llaser = 594\,$nm illumination light, $\Ton \sim 2\,$s long bursts of $n\sim 600$ photons become visible separated by $\Toff \sim 18\,$s of background fluorescence.
\\
\indent
The demonstrated optically induced charge state dynamics and fluorescence response are homogeneous for all NV centers in bulk diamond.
This is a valuable property for a scalable parallel superresolution microscopy technique.

\subsection{STORM with NV centers in diamond}      
\begin{figure}[t]
  \centerline{\includegraphics[width=1.0\columnwidth]{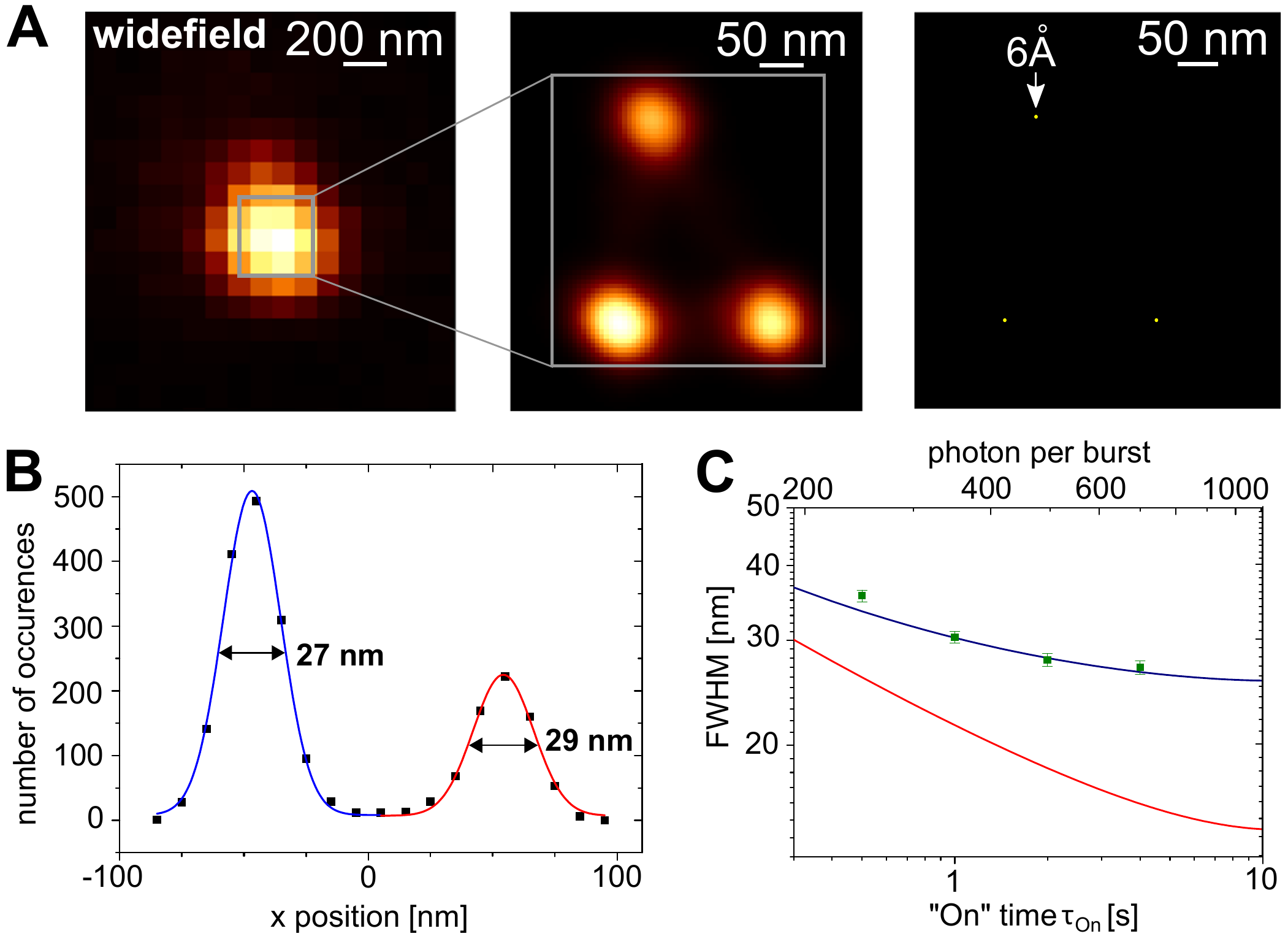}}
  \caption{\textbf{Sub diffraction-limit imaging of NV centers by STORM.}
	(\textit{A}) From left to right: Diffraction-limited CCD image of three unresolved NV centers.
	STORM image of the formerly unresolved centers shows three distinct location distributions.
	Emitter localization accuracy is further reduced due to multiple photon bursts per NV.
	The STORM settings were $1\,$s average ``On'' time and on average $350\,$photons per burst.
	(\textit{B}) 1d line scan through a STORM image.
	The emitter location distribution (x-coordinate) resolves two NV defects with $\mathrm{FWHM} = 27\,$nm respectively $29\,$nm.
	Here the average ``On'' time was $4\,$s.
	(\textit{C}) Resolution as a function of the ``on'' time or the number of photons per burst, respectively.
	Green squares are measurement data, the blue curve is a fit according to eqs.~(\ref{eq:pos_accuracy_thompson}),(\ref{eq:pos_accuracy_thompson_drift}), considering the changing number of signal and background photons with increasing ``on'' time.
	The red curve shows the theoretical resolution, when the contribution caused by sample drift is neglected.
	\label{fig:storm}}
\end{figure}

We demonstrate STORM on three NV centers within a diffraction limited spot.
In fig.~\ref{fig:storm}\textit{A} we compare the resulting images from conventional (left) and STORM (center) imaging.
For conventional imaging we illuminate emitters with $532\,$nm laser light at saturating power levels (i.e. with laser intensity $\Ilaser \approx 200\,$kW/cm$^2$) resulting in a CCD image exhibiting a single fluorescent spot.
For STORM imaging, in contrast, we apply $594\,$nm laser light with intensities on the order of $\Ilaser \approx 1\,$kW/cm$^2$ (i.e. far below saturation) finally yielding a reconstructed image showing three distinct NV emitters.
\\
\indent
When switching to the low intensity $594\,$nm laser light for STORM, we start seeing distinguishable fluorescence photon bursts (see fig.~\ref{fig:setup}\textit{C}) either on a single photon counting module (APD) or a CCD camera.
\\
\indent
For STORM imaging we record CCD images at a constant rate.
As the ``On''-''Off''-switching of fluorescence happens stochastically, it is therefore not synchronized with the CCD frames.
We set the exposure time of the CCD camera to the average ``On''-time $\Ton$ of the emitters.
As our emitters do not bleach we can record many bursts per NV center.
Finally, the asynchronous switching of the emitters necessitates post processing of the data (see methods).
\\
\indent
The idea of STORM is to assign all photons $n_i$ of a single, localized burst $i$ on the CCD to a single yet unknown emitter.
To this end, all photons $n_i$ are used to calculate an average center location $[x_i,y_i]$ with a reduced location uncertainty $\sigma_{x/y,i}$ as compared to the diffraction limit $\sigma_{\lambda,i}$.
The improved uncertainty scales approximately as $\sigma_{x/y,i} \propto \sigma_{\lambda,i}/\sqrt{n_i}$ (see Methods).
Eventually, summing up Gaussian location distributions with parameters $[x_i,y_i]$ and $\sigma_{x/y,i}$ for all photon bursts yields fig.~\ref{fig:storm}\textit{A} (center) where three individual emitters are clearly distinguishable.
The FWHM of the location distribution is $27\,$nm (see fig.~\ref{fig:storm}\textit{B}) and in the absence of drift is projected to be $\approx 14\,$nm (see fig.~\ref{fig:storm}\textit{C} and Methods).
In the end, each obtained emitter location corresponds to a particular, distinguishable subset of distributions (each of the three spots in fig.~\ref{fig:storm}\textit{A} (center)).
These distributions can be used to further improve the localization accuracy of that particular emitter to $6\,$\AA\ in the present case (see fig.~\ref{fig:storm}\textit{A} (right) and Methods).

\subsection{Sub diffraction-limit magnetic resonance}
\begin{figure}[t]
  \centerline{\includegraphics[width=1.0\columnwidth]{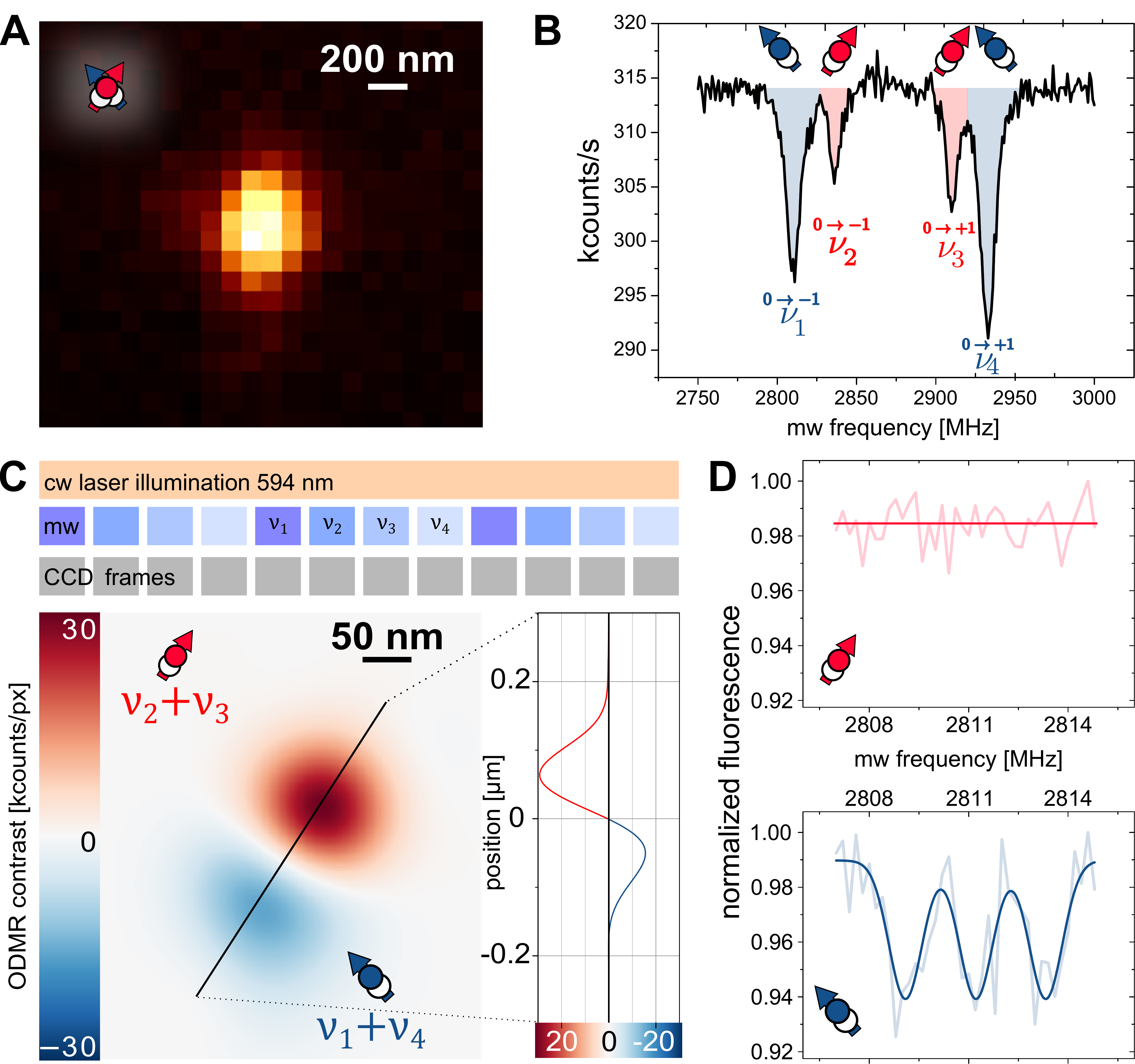}}
  \caption{\textbf{nano-scale magnetic field sensing with NV defects.}
	(\textit{A}) Diffraction limited microscopy image of two NV centers with different crystallographic orientations (see inset sketch).
	(\textit{B}) Conventional cw ODMR of NV pair exhibiting two pairs of resonance lines.
	Addressing of individual spins is possible due to different NV orientation but assignment to a location is missing.
	(\textit{C}) (top) STORM-ODMR measurement sequence (see text).
	The average NV ``On'' time was $1\,$s.
	(bottom) STORM-ODMR image (see text).
	The zero crossing (see linescan) between both centers will even appear for mutual distances below STORM resolution.
	(\textit{D}) High resolution STORM-ODMR spectra assigned to individual NV centers.
	The mw frequencies are centered around $\nu_1$.
	Thus, only the blue NV shows a resonance.%
  \label{fig:odmr}}
\end{figure}

As the photon count rate during fluorescence bursts of STORM imaging does depend on the electron spin state,
tagging of photons by the NV spin state is feasible.
An exemplary \textit{conventional} ODMR spectrum (see Methods) of two spatially unresolved NV centers (see fig.~\ref{fig:odmr}\textit{A}) is shown in fig.~\ref{fig:odmr}\textit{B} where the outer spectral lines ($\nu_1, \nu_4$) belong to one NV and the inner resonances ($\nu_2,\nu_3$) to the other one.
Using \textit{conventional} ODMR, however, no assignment of spatial to spectral information is possible.
In contrast, STORM \textit{in conjunction} with ODMR (STORM-ODMR) allows tagging the spatial location of NV centers with spectral information to achieve said assignment (see fig.~\ref{fig:odmr}\textit{C}). 
\\
\indent
In order to apply STORM-ODMR, we perform STORM as described while simultaneously applying mw radiation and switching its frequency repeatedly from $\nu_1$ through $\nu_4$ synchronously to the CCD frames. 
The resulting image is shown in fig.~\ref{fig:odmr}\textit{C}. 
Here, we have added up all location distributions of photon bursts for mw frequencies $\nu_{1,4}$ and subtracted those for $\nu_{2,3}$.
As a result we see a red (positive) and a blue (negative) distribution of locations, belonging to the NV with resonances $\nu_{2,3}$ and the one with resonances $\nu_{1,4}$ respectively.
Both distributions are separated by a zero crossing (see linescan in fig.~\ref{fig:odmr}\textit{C}).
Please note that this zero crossing will occur for every distance of the two emitters even if they are closer than the STORM resolution $\sigma_{x/y,i}$.
This shows that ODMR can be used to enhance the STORM resolution (see Methods) similar to diffraction limited microscopy like demonstrated in \cite{dolde_room-temperature_2013,maurer_far-field_2010,grinolds_nanoscale_2013,chen_wide-field_2013}.
\\
\indent
Further on we show that localization is not only sensitive to the electron but also to nuclear spin states.
Utilizing \textit{high spectral resolution} magnetic resonance enables to reveal hyperfine coupling to proximal nuclear spins.
To this end, we reduce the mw power to avoid a related broadening of the ODMR resonance lines, and thus to exploit the small electron spin relaxation rate.
The latter is mainly limited by the $^{13}$C nuclear spin bath \cite{mizuochi_coherence_2009}.
We demonstrate \textit{high spectral resolution} by sampling the frequency range around resonance position $\nu_1$ obtained in the previous ODMR measurement.
Consequently, we are able to assign a partial high resolution ODMR spectrum to each individual NV center (see fig.~\ref{fig:odmr}\textit{D}).
As expected, one NV center shows ODMR resonances in this spectral window whereas the other one does not.
In the corresponding spectrum we can resolve the hyperfine coupling to the adjacent $^{14}$N nuclear spin.
The summed up contrast of the STORM-ODMR spectrum of $18\,\%$ is comparable to that of conventional ODMR spectra on NV centers (i.e. $\approx 30\,\%$).
To further discriminate individual, proximal bath spins dynamical decoupling sequences need to be applied \cite{zhao_sensing_2012}.
\\
\indent
From the resonance lines' slopes and contrasts in the STORM-ODMR spectrum in fig.~\ref{fig:odmr}\textit{D} we estimate the magnetic field sensitivity for a single NV spin to be $\delta B \approx 190\, \mu \mathrm{T /\sqrt{Hz}}$ (see Methods).
Combined with the advantage of parallel imaging we can in principle measure the magnetic field at all accessible centers at once.

\section{Discussion}
Summarizing we demonstrated the first optical superresolution imaging technique with parallel data acquisition for NV centers in bulk diamond.
Additionally, we were able to combine superresolution imaging with spin resonance techniques.
Due to the homogeneity of the exploited properties among NV centers in bulk diamond our method is intrinsically scalable to a vast number of color centers simultaneously.
\\
\indent
Regarding the optical super-resolution imaging of NV centers, we developed a dedicated STORM technique achieving a FWHM resolution of single emitters of $27\,$nm and a localization accuracy of $6\,$\AA .
As our emitters inherently do not bleach or move with respect to the diamond itself, it is a potential calibration sample for STORM microscopes or a fixed background reference for experiments on otherwise moving objects of study.
Furthermore, there is no need for high intensity or even pulsed lasers as illumination source which makes our method extremely valuable for biological experiments \textit{in vivo} or any other light or temperature sensitive measurements.
\\
\indent
Regarding nanometer-scale spin resonance, we combined STORM with ODMR to demonstrate NV electron spin readout with a spatial resolution far below the diffraction limit.
We have utilized this technique to further improve spatial resolution of the imaging technique and for highly localized nuclear spin detection.
It is therefore particularly valuable for applications of diamond as an "microscope sensor slide". Here, dense ensembles of NV centers (i.e. distances of $\sim 10\,$nm) are placed close ($\sim 1\,$nm) to the diamond surface and can then detect physical quantities (e.g. magnetic and electric fields or nuclear spin concentrations) originating from samples just outside of the diamond \cite{steinert_high_2010,steinert_magnetic_2013,le_sage_optical_2013,maertz_vector_2010,pham_magnetic_2011,hall_high_2012, staudacher_nuclear_2013,mamin_nanoscale_2013}.
\\
\indent
Another application of STORM with NV centers could be fluorescent nano-diamonds used as bio markers.
At the current time, application of our STORM technique to nano-diamonds is challenging due to the large inhomogeneity of NV and nanodiamond properties.
For example, fluorescence intensities, charge state ratios as well as timescales and spectral response of charge state dynamics differ greatly among different NV centers and nanodiamonds. 
The latter observation was attributed to electron tunneling and differing Fermi levels among nanodiamonds in \cite{gu_super-resolving_2013}.
Nevertheless, in \cite{gu_super-resolving_2013}, a similar charge state switching mechanism was used to superresolve NV centers.
\\
\indent
Beyond our proof-of-principle experiments there is room for further improvement.
With the reduction of sample drift and the increase of photon collection efficiency a FWHM resolution below $\approx 10\,$nm is achievable in the short term.
In the longer term, control of the Fermi level, diamond doping and surface preparation \cite{hauf_chemical_2011,grotz_charge_2012} might allow for a wider range of possible illumination intensities and wavelengths and thus higher acquisition speeds and a tailored degree of parallelism.
Furthermore, we have sketched a road towards a fully parallel two-dimensional quantum sensor array with nanometer scale resolution.
Some of its features would be for example large area parallel magnetic field sensing which outperforms scanning techniques by orders of magnitude (see methods), or the direct imaging of spatially distributed quantum correlations on length scales down to nanometers.

\appendix
\section{Experimental setup} The setup can be divided into two main parts which are the optical microscope to address single NV centers and the spin manipulation equipment consisting of microwave (mw) sources and static magnetic fields (see fig.~\ref{fig:setup}\textit{A}).
The microscope is capable of both wide-field CCD imaging and scanning confocal imaging.
For illumination we use lasers with wavelengths 532$\,$nm and 594$\,$nm which can be both switched on the order of ns and which are intensity controlled.
In order to detect fluorescence only from negatively charged NV centers an optical longpass filter with a cutoff wavelength of 650$\,$nm is used.
For adjusting the electron spin energy eigenstates and the respective transition frequencies we place and orient permanent magnets accordingly.
Microwave radiation for spin manipulation is guided via copper wires close to the NV centers.
\\
\indent
For confocal imaging a collimated beam is sent into an high NA oil immersion objective and fluorescence light is collected by the same objective and finally focused onto a pinhole for spatial filtering and then onto a single photon counting avalanche photo diode (APD, see fig.~\ref{fig:setup}).
A 3D piezostage with nanometer precision is used to move the objective across the diamond sample.
For CCD imaging the piezoscanner is fixed and additional lenses can be flipped into the beam path to focus the illumination laser onto the back focal plane of the objective which leads to wide-field illumination.
In that case the fluorescence light is guided onto the pixel array of an electron multiplying CCD camera (iXon Ultra 897, Andor Technology).
The magnification is set such that $100\,$nm in the focal plane corresponds to one pixel on the CCD (see fig.~\ref{fig:setup}\textit{A}).
\\
\indent
We use commercial CVD diamond samples of type IIa which contain as grown and artificially created NV centers.

\section{NV charge state switching for STORM}
For illumination with $\Llaser$=594$\,$nm light and intensities $\Ilaser$ far below saturating levels, the NV center fluorescence level exhibits sudden jumps which can be attributed to charge state switching from NV$^-$ to NV$^0$ induced by two photon ionization (see fig.~\ref{fig:setup}\textit{C}) \cite{waldherr_dark_2011,aslam_photo-induced_2013}.
This behavior can be characterized and controlled as follows.
The average ``On'' and ``Off'' times, $\Ton$ and $\Toff$, of fluorescence bursts and background fluorescence respectively are inversely proportional to $\Ilaser^2$ ($\Ton,\Toff \propto \Ilaser^{-2}$).
In addition, The fluorescence count rate $\Gamma$ during a burst is directly proportional to $\Ilaser$ ($\Gamma \propto \Ilaser$).
Consequently, the average number of photons per burst $n$ is proportional to the square-root of $\Ton$ and inversely proportional to $\Ilaser$ ($n\propto\sqrt{\Ton},\, n\propto \Ilaser^{-1}$; see fig.~\ref{fig:setup}\textit{D}).
The ratio of $\Ton$ and $\Toff$ can be altered by tuning the illumination wavelength (see fig.~\ref{fig:setup}\textit{D}).
For 594$\,$nm illumination light the $\Ton$ fraction is $r\approx 0.1$, and it decreases for longer wavelengths.

\section{STORM}
During STORM imaging we continuously record CCD images at a constant rate while the ``On''-''Off''-switching of fluorescence happens stochastically.
This necessitates post processing of the data which is divided into two steps.
First, we select frames with just a single active emitter by setting upper and lower thresholds for the photon number.
Thus we discard empty frames or those with two or more active emitters.
In a second step we check for lateral asymmetries in the fitted PSF of all remaining frames as an additional sign for more than one active emitter.
\\
\indent
For the remaining frames a 2d Gaussian function is fitted to frame number $i$ which yields a center location $[ x_{i} \left( n_{i} \right),y_{i} \left( n_{i} \right) ]$, the number of photons $n_i$ and the width $\sigma_{\lambda,i}$ of the point spread function.
Please note that the Gaussian function is just an approximation of the real point spread function (PSF) \cite{webb_confocal_1996}.
The width $\sigma_{\lambda,i}$ of the fitted PSF allows to determine the relative axial location of the emitter with respect to the focal plane.
In our case however all emitters are in the same focal plane.
As the number of photons contributing to the center location is $n_i$ the expected standard deviation of that location should scale as $\sigma_{x/y,i} \propto \sigma_{\lambda,i}/\sqrt{n_i}$.
Sensor pixelation and background noise, however, affect the standard deviation like \cite{thompson_precise_2002}
\begin{equation}
\sigma_{x/y,i}^{2} = \frac{\sigma_{\lambda,i}^{2}}{n_i}  + \frac{a^{2}/12}{n_i} + \frac{8 \pi \sigma_{\lambda,i}^{4} b^{2}}{a^{2} \cdot n_i^{2}}
\label{eq:pos_accuracy_thompson}
\end{equation}
where $a$=100$\,$nm is the pixel size of the images and $b(t)\approx 1+0.3\,\mathrm{s}^{-1} \cdot t$ accounts for background photons per pixel in dependence of exposure time $t$.
In our STORM technique the average number of photons $n_i$ increases with the square-root of the burst length/ exposure time $t$, where the burst length is altered by the illumination intensity.
Hence, we expect an optimal time for minimal standard deviation $\sigma_{x/y,i}^{2}$ according to eq.~(\ref{eq:pos_accuracy_thompson}) (see fig.~\ref{fig:storm}\textit{C}).
\\
\indent
With a set of locations $[x_i,y_i]$ and standard deviations $\sigma_{x/y,i}$ for each valid CCD frame we construct a new emitter location distribution
\begin{equation}
p(x,y) = \sum_i{\frac{1}{2\pi\sigma_{x/y,i}^2} \exp{\left( -\frac{(x-x_i)^2+(y-y_i)^2}{2\sigma_{x/y,i}^2}\right)}}
\label{eq:new_position}
\end{equation}
by adding up 2d Gaussians with the respective parameters and unit weight.
Figure \ref{fig:storm}\textit{B} and \textit{C} shows the resulting super-resolution image with typical FWHM of the location distributions of $\approx 28\,$nm.
\\
\indent
For on average $\approx 700$ photons per burst and average burst lengths of 4$\,$s these location distributions are broader than what is expected from eq.~(\ref{eq:pos_accuracy_thompson}).
This is due to sample drift during the accumulation of CCD frames which leads to additional broadening when all bursts are summed up according to eq.~(\ref{eq:new_position}).
Thus, the overall standard deviations of the emitter locations can be estimated as
\begin{equation}
\sigma_{x/y}^{2} = \sigma_{x/y,i}^{2} + \sigma_{\mathrm{drift}}^{2}
\label{eq:pos_accuracy_thompson_drift}
\end{equation}
where we have evaluated the drift to be $\sigma_{\mathrm{drift}} \approx 10\,$nm.
\\
\indent
As the achieved standard deviation $\sigma_{x/y}\approx 28\,$nm is smaller than the average distance between the emitters we are able to recognize single distinct location distributions for individual NV centers.
Thus, we can compute the localization accuracy $\hat{\sigma}_{x/y}=\sigma_{x/y}/\sqrt{M}$ where we ideally gain an additional factor of $1/\sqrt{M}$ where $M$ is the number of detected bursts for an individual NV center (see fig.\ref{fig:storm}\textit{B}).
With an average number of bursts per NV of $\approx 2000$ we achieve a localization accuracy of $\hat{\sigma}_{x/y}$=6$\,$\AA .
\\
\indent
The presented resolution in fig.~\ref{fig:storm}\textit{C} was achieved under optimal conditions (i.e. optimal laser intensity) with respect to standard deviation $\sigma_{x/y}$ of location distributions.
The optimum of $\sigma_{x/y}$ arises from an increase of $n\propto \sqrt{\Ton}$ for decreasing laser intensity (i.e. increasing exposure time $\Ton$) on the one hand and an increase in background noise $b\propto \Ton$ on the other hand (see eq.~\ref{eq:pos_accuracy_thompson}).
For laser powers below the optimum value, $b$ will become the leading term resulting in worsening resolution.
In fig.~\ref{fig:storm}\textit{C} the FWHM of the location distribution is presented for different average burst durations (i.e. different laser powers).
The theoretical estimation is in agreement with these values.
We like to stress that the current resolution is mainly limited by sample drift.
Accordingly, we estimate FWHM in the absence of sample drift to be $\approx 14\,$nm.

\section{Sub diffraction-limit magnetic resonance}
At first conventional ODMR is performed on two NV centers within one diffraction-limited spot.
To this end 532$\,$nm laser light (with intensities saturating the optical transition of the NV centers) and mw radiation (capable of spin transition rates of $\approx 10\,$MHz) are continuously on, the mw frequency is swept and the corresponding fluorescence is recorded.
The laser pumps the NV spin into projection $m_S$=0 whereas resonant mw radiation induces transitions $m_S=0 \leftrightarrow \pm1$ (see fig.~\ref{fig:setup}\textit{B}).
As the Zeeman interaction splits the $m_S=\pm1$ levels there are usually two different resonances.
Away from spin resonance the spin state is $m_S=0$ and accordingly a high rate of fluorescence photons is obtained.
Upon resonance conditions the levels $m_S=\pm1$ become populated and the fluorescence decreases.
Although both NV centers are exposed to the same external magnetic field, by chance their symmetry axes lie along different directions of $\left\langle 111 \right\rangle$ in the diamond lattice resulting in crystal fields with different directions \cite{balasubramanian_nanoscale_2008}.
Thus the vectorial sum of external and crystal field is different for both NV centers and thus are their ODMR resonance lines.
\\
\indent
Fig.~\ref{fig:odmr}\textit{C} demonstrates mw enhancement of STORM resolution.
The zero crossing of the linescan in fig.~\ref{fig:odmr}\textit{C} will be visible for any distance $d$ between two NV centers.
However, the depth of the minimum and maximum will approximately decrease proportional to $d$.
Thus, to achieve a unit signal to noise ratio the number of accumulated photons for decreasing $d$ goes as $1/d^{2}$ or inverse measurement time squared.
In other words, the resolution increases further as square root of the photon number. 
\\
\indent
For high spectral resolution magnetic resonance we have reduced mw power to avoid power broadening of the resonance lines and we have sampled the resonance line $\nu_1$ in small frequency steps.
For each valid CCD frame we noted the corresponding mw frequency.
As in the previous measurement, one NV shows a resonance around $\nu_1$ and the other one does not.
Apparently, the linewidth is drastically reduced and the hyperfine interaction of 2.2$\,$MHz to the nitrogen nuclear spin with total nuclear spin $I=1$ is visible.
The current linewidth limit is set by the $^{13}$C nuclear spin bath \cite{mizuochi_coherence_2009}.
\\
\indent
From the high resolution STORM-ODMR we can calculate the magnetic field sensitivity.
Therefore we can take into account all three hyperfine lines ($m_I=-1,0,1$) because they would be shifted commonly upon a change in magnetic field.
For the sensitivity we arrive at
\begin{equation}
\delta B = \left( \sum_{m_I=-1..1}{ \left. \frac{d\Gamma}{d\nu} \right|_{\mathrm{max}} } \right)^{-1} \sigma_{\Gamma} \frac{2\pi}{\gamma_e} \sqrt{T}
\label{eq:sensitivity}
\end{equation}
where $d\Gamma/d\nu_{\mathrm{max}}$ is the maximum slope at each hyperfine line, $\sigma_{\Gamma}$ is the standard deviation of the data from the fit, $\gamma_e$ is the gyromagnetic ratio of the electron spin and $T$ is the total measurement time.
Finally, the achieved sensitivity is $\delta B \approx 190\, \mu \mathrm{T /\sqrt{Hz}}$.
This value is four times higher than what is expected for bare photon shot noise limitations.
We attribute this mismatch to the post processing of the CCD frames and fluctuating laser intensity.
\\
\indent
The magnetic field sensitivity for conventional single NV experiments are better because of the higher fluorescence count rate.
Under similar conditions (i.e. same ODMR linewidth) an ideal single NV experiment (shot noise limited) would yield $\approx 180\,$nT/$\sqrt{\mathrm{Hz}}$ \cite{taylor_high-sensitivity_2008}.
Consequently, the conventional sensing technique is $\approx 10^6$ times faster for a single NV center.
However, when scanning many NV centers in parallel and with nanometer spatial resolution this will eventually pay off.
For example, with a field of view of $100 \times 100\,\mu$m$^2$ and CCD with $1024\times1024$ pixels we can resolve and consequently measure the local magnetic field of $\approx 10^7$ NV centers simultaneously given a FWHM spatial resolution of $\approx 30\,$nm.
A scanning super-resolution technique like STED ideally might achieve the mentioned sensitivity of $\delta B \approx 180\,$nT$\sqrt{\mathrm{Hz}}$.
However, as the number of NV centers in our example is $10^7$ its overall speed is still one order of magnitude slower than our STORM-ODMR method.
With foreseeable improvements the speed of STORM-ODMR can be increased by many orders of magnitude (see below).

\section{STORM-ODMR prospects}
Our novel technique of using STORM in conjunction with ODMR on NV centers can be used for building very powerful quantum sensors.
The latter exploits the sensitivity of the NV center's spin transition on quantities like magnetic and electric fields, strain or temperature.
For the first time a truly parallel quantum sensor array with nanometer scale ``pixel'' size can be envisioned.
Some of its features would be for example large area parallel magnetic field sensing or the direct imaging of spatially distributed quantum correlations on length scales down to nanometers.
\\
\indent
Taking into account current NV diamond properties we are going to estimate the achievable sensitivity of a potential NV sensor based on STORM-ODMR.
With respect to magnetic field sensitivity it is optimal to have the smallest possible ODMR linewidth.
For full benefit with cw laser illumination, the optical excitation rate must be similar to the decoherence rate (i.e. inverse linewidth).
In our demonstrated nano-scale ODMR the linewidth was limited to $\approx 1\,$MHz by the diamond nuclear spin bath \cite{mizuochi_coherence_2009}.
For comparison, the optical excitation rate was $\sim 1\,$kHz.
Thus, a potential reduction of the ODMR linewidth down to the order of the optical excitation rate (as demonstrated in \cite{mizuochi_coherence_2009}) would yield a linear sensitivity enhancement of three orders of magnitude.
For further improvement in sensitivity both, optical excitation rate and magnetic resonance linewidth, would have to be decreased equally and the sensitivity improvement would then scale as the square root of the rate reduction \cite{taylor_high-sensitivity_2008}.
\\
\indent
In addition to rather incoherent cw ODMR techniques also pulsed schemes are applicable.
These allow for a higher versatility owing to coherent spin control \cite{balasubramanian_ultralong_2009,staudacher_nuclear_2013,dolde_electric-field_2011,neumann_high-precision_2013}.
To this end, laser and mw control of the spins are interleaved \cite{balasubramanian_ultralong_2009} to prevent optical excitation during coherent spin operations.
\\
\indent
Summarizing, STORM-ODMR enables the same measurement possibilities as demonstrated for single NV centers with the benefit of an increased spatial resolution and highly parallel control and readout, however, with a lower fluorescence count rate per emitter.
The latter drawback can be mitigated in ultrahigh sensitivity metrology applications where spin transitions with homogeneous linewidths of $\sim 1\,$kHz are exploited \cite{balasubramanian_ultralong_2009}.
In these cases average fluorescence count rates for conventional single NV measurements and STORM-ODMR measurements approach the order of emitter ``On'' and ``Off'' time ratio which is $\approx 10$ in our experiment.
Thus conventional measurements would be $\sim 10$ times faster for a single NV center.
Eventually, parallel STORM-ODMR measurements on as few as $\sim 10$ NV nanoprobes would perform equally well as conventional serial measurements on $\sim 10$ single NV centers.
With a $100\times100\,\mu$m$^2$ area, $60\,$mW laser power, a FWHM resolution of 14$\,$nm and a corresponding NV density, parallel measurements would be up to $10^6$ times faster than serial super-resolution measurements.
\\
\indent
The currently sketched sensor array does not operate fully parallel, which means NV centers within a confocal spot are readout in serial.
To take an instant snapshot of the whole sensor array with nanometer resolution, nuclear spins can be employed as non-volatile memory \cite{neumann_single-shot_2010,maurer_room-temperature_2012}.
To this end, all NV center probes sense at the same time \cite{steinert_high_2010}, their results are stored on their individual proximal nuclear spins (e.g. $^{14}$N, \cite{neumann_single-shot_2010}) which are then read out via STORM-ODMR.
This way even delocalized quantum correlations can be monitored.

\begin{acknowledgments}
Thanks to Steffen Steinert, Torsten Rendler and Petr Siyushev for assistance with the wide-field microscopy setup.
We also acknowledge Andrej Denisenko for bulk diamond sample preparation and Markus Remm improving mechanical stability of the setup.
We thank Michael B\"orsch for valuable discussions.
We thank L.O.T-QuantumDesign GmbH for lending us the iXon Ultra 987 EMCCD camera.
Our research is supported by the EU via SQUTEC and Diamant, by the DFG via the SFB/TR21 and research group 1493 “Diamond quantum materials”, and by Baden-W\"urttemberg Stiftung gGmbH (“Methoden f\"ur die Lebenswissenschaften”).
\end{acknowledgments}

\bibliography{}

\end{document}